\documentstyle[12pt,epsfig]{article}
\begin{document}

\begin{center}
{\Large \bf Semileptonic decays of the bottom-charm hadrons} \\

\vspace{4mm}

M.A. Ivanov\\

\vspace{4mm}
Bogoliubov Laboratory of Theoretical Physics, \\
Joint Institute for Nuclear Research\\
141980 Dubna, Russia\\
\end{center}

\begin{abstract}
We present the results of a study of the semileptonic decays of the 
$B_c$-meson and the lowest lying doubly heavy baryons 
using the relativistic quark model.  We do not employ a heavy quark
mass expansion but keep the masses of the heavy quarks and hadronss finite.
We calculate all relevant form factors and decay rates.  
\end{abstract}

The  semileptonic decays of heavy mesons and baryons are ideally suited 
to extract the Cabibbo-Kobayashi-Maskawa (CKM) matrix elements.
The heaviest flavored bottom-charm $B_c$-meson was observed by the 
CDF Collaboration \cite{CDF} in the analysis of the decay mode 
$B_c\to J/\psi \bar l \nu$. The discovery of the $B_c$-meson raises hopes 
that doubly heavy flavored baryons will also be discovered in the near future.
The theoretical treatment of the systems with two heavy quarks is complicated 
by the fact that one cannot make use of an expansion in terms of the inverse 
heavy quark masses. Previously, nonrelativistic potential models, diquark 
approximation, QCD sum rules and nonrelativistic QCD have been used to 
describe the  spectroscopy of doubly heavy baryons and to estimate the 
inclusive and some exclusive decay modes of such systems (for review, 
see \cite{Gersh}-\cite{GMS} and references therein).

We present here our recent results \cite{Bc,dhb} of exploration of
the semileptonic decays of the doubly heavy $B_c$-meson and 
the lowest lying doubly heavy $\Xi$-baryons within a relativistic 
constituent quark model. This model \cite{model}  can be viewed as 
an effective quantum field theory approach based on an  interaction 
Lagrangian of hadrons  
interacting with their constituent quarks. Universal and reliable predictions 
for exclusive processes involving both  mesons composed from a quark and 
antiquark and baryons composed from three quarks result from this approach. 
The coupling strength of hadrons $H$  to their constituent quarks
is determined by the compositeness condition $Z_H=0$ \cite{SWH,EI} where 
$Z_H$ is the wave function renormalization constant of the hadron. The 
quantity $Z_H^{1/2}$ is the matrix element between a physical particle state 
and the corresponding bare state. The compositeness condition $Z_H=0$ enables 
us to represent a bound state by introducing a hadronic field interacting 
with its constituents so that the renormalization factor is equal to zero. 
This does not mean that we can solve the QCD bound state equations but we are 
able to show that the condition $Z_H=0$ provides an effective and 
self-consistent way to describe the coupling of the particle to its 
constituents. One  starts with an effective interaction Lagrangian written 
down in terms of quark and hadron variables. Then, by using Feynman rules, 
the $S$-matrix elements describing hadron-hadron interactions are given in 
terms of a set of quark diagrams. In particular, the compositeness condition 
enables one to avoid the double counting of quark and hadron degrees of 
freedom. This approach is self-consistent and all calculations of physical
observables are straightforward. There is  a small set of  model parameters: 
the values of the constituent quark masses and the scale parameters that 
define the size of the distribution of the constituent quarks inside a given 
hadron. The shapes of the vertex functions and the  quark  propagators can in 
principle  be determined from an analysis of the Bethe-Salpeter (Fadde'ev) 
and Dyson-Schwinger equations, respectively, as done e.g. in \cite{DSE,DSEH}. 
In the present paper we, however, choose a more phenomenological approach 
where the vertex functions are modelled by Gaussian forms and the quark 
propagators are given by local representations. We have demonstrated in our 
papers \cite{RTQM} that the relativistic constituent model is consistent 
with the heavy quark symmetry in the limit of infinite quark masses.

We start with the effective interaction Lagrangian which describes 
the coupling between
hadrons and their constituent quarks. For example, the couplings of
the bottom-charm hadrons into their  constituents
are given by  

\begin{eqnarray}
{\cal L}_{{\rm int}}(x)&=&g_{B_c}\, B_c(x)\,J_{B_c}(x)+
   \left(g_{\Xi_{bc}}\,\bar\Xi_{bc}(x)\,J_{\Xi_{bc}}(x)+{\rm h.c.} \right)\,,
\label{lag}\\
&&\nonumber\\
J_{B_c}(x)&=&\int\!\! dx_1 \!\!\int\!\! dx_2
\Phi_{B_c} (x;x_1,x_2)\, \left(\bar b(x_1)i\gamma^5  c(x_2)\right)\,,
\nonumber\\
&&\nonumber\\
J_{\Xi_{bc}}(x)&=&
\int\!dx_1\!\int\!dx_2\!\int\!dx_3\,
\Phi_{\Xi_{bc}}(x;x_1,x_2,x_3)\,
\gamma^\mu\gamma^5 b_{a_1}(x_1) 
\left(c_{a_2}(x_2)\,C\gamma^\mu\, q_{a_3}(x_3)\right) 
\varepsilon^{a_1a_2a_3} \,, \nonumber
\end{eqnarray}
where $q=u,d$. The vertex function $\Phi_H$ is taken to be invariant under 
the translation $x\to x+a$ which guarantees Lorentz invariance for the 
interaction Lagrangian Eq.~(\ref{lag}).

The matrix elements of the weak current $\bar b O^\mu c$ are written
in our approach as

\begin{eqnarray}
<\eta_c(p^{\,\prime})|\bar b O^\mu c|B_c(p)>&=&
3g_{B_c}g_{\eta_c}\!\!\int\!\!\frac{d^4k}{(2\pi)^4i} 
\tilde\Phi_{B_c}(-k^2)\tilde\Phi_{\eta_c}(-k^2)
\label{eta}\\
&& 
\cdot
{\rm tr}\biggl[S_c(\not\! k+\not\! p^{\,\prime}) O^\mu S_b(\not\! k+\not\! p)
\gamma^5 S_c(\not\! k) \gamma^5 \biggr]
\nonumber\\
&&\nonumber\\
&=&f_+(q^2)\;(p+p^{\,\prime})^\mu\,+\,f_-(q^2)\;(p-p^{\,\prime})^\mu\,,
\nonumber\\
&&\nonumber\\
<J/\psi(p^{\,\prime}),\epsilon^{\ast}|\bar b O^\mu c|B_c(p)>&=&
3g_{B_c}g_{J/\psi}\!\!\int\!\!\frac{d^4k}{(2\pi)^4i} 
\tilde\Phi_{B_c}(-k^2)\tilde\Phi_{J/\psi}(-k^2)
\label{psi} \\
&&
\cdot
{\rm tr}\biggl[S_c(\not\! k+\not\! p^{\,\prime} ) O^\mu S_b(\not\! k+\not\! p)
\gamma^5 S_c(\not\! k) \not\! \epsilon^{\ast}\biggr]
\nonumber\\
&=& 
 -\epsilon^{\ast\mu}
(m_{B_c}+m_{J/\psi})\,A_1(q^2)+ 
(p+p^{\,\prime} )^\mu\,  p\cdot \epsilon^\ast \,
\frac{A_2(q^2)}{m_{B_c}+m_{J/\psi}}+
\nonumber\\
&& +(p-p^{\,\prime})^\mu\, p\cdot \epsilon^\ast \,\frac{A_3(q^2)}{m_P+m_V}
-i\varepsilon ^{\mu\nu\alpha\beta}\epsilon^{\ast\nu} p^\alpha
p^{\,\prime~\beta }\, \frac{2\ V(q^2)}{m_P+m_V}\,, 
\nonumber\\
&&\nonumber\\ 
<\Xi_{cc}(p^{\,\prime} )|\bar b O^\mu c|\Xi_{bc}(p)>&=&
12g_{\Xi_{cc}}g_{\Xi_{bc}}
\int\!\frac{d^4k_1}{(2\pi)^4i}\!\int\!\frac{d^4k_2}{(2\pi)^4i}\,
\tilde\Phi_{\Xi_{cc}}(-K^2)\,\tilde\Phi_{\Xi_{bc}}(-K^2)
\nonumber\\
&&
\cdot\gamma^\alpha\gamma^5\,S_q(k_2)\,
\gamma^\beta\,S_c(k_1+k_2)\,\gamma^\alpha\, S_c(k_1+p^{\,\prime}) \, 
O^\mu\, S_b(k_1+p)\,\gamma^\beta\gamma^5 \,, \nonumber\\
&&\nonumber\\ 
&=&
\gamma^\mu\,(F_1^V-F_1^A\,\gamma^5)\,+\, i\sigma^{\mu\nu}q^\nu\, 
(F_2^V-F_2^A\,\gamma^5)\,+\, q^\mu\,(F_3^V-F_3^A\,\gamma^5)\,\nonumber
\end{eqnarray}
where $K^2\equiv k_1^2+(k_1+k_2)^2+k_2^2$. 
The quark propagator is chosen to have a local form
\begin{equation}\label{prop}
S_i(k)=\frac{1}{m_i-\not\! k} \hspace{1cm}
(i=u,d,s,c,b)
\end{equation}
with $m_i$ being a constituent quark mass.
We assume that $m_H<\sum_{i=1}^n m_{q_i}$ 
(n=2 for mesons and n=3 for baryons) in order to
avoid the appearance of imaginary parts in the physical
amplitudes. This is a reliable approximation for the heavy
pseudoscalar mesons. The above condition is not always met for heavy vector
mesons. We shall therefore employ equal masses for the
heavy pseudoscalar and vector mesons in our form factor calculations
but use physical masses for the phase space.
 
Generally, $\tilde\Phi_H$ is a function of extrernal  momenta too, 
however, in the impulse approximation employed in our approach, 
we assume that it only depends on the sum of relative momentum 
squared.

The coupling constants $g_H$ are determined  by the so called {\it
compositeness condition} proposed in \cite{SWH} and extensively
used in \cite{EI}. The compositeness condition means that the
renormalization constant of the meson field is equal to zero
$Z_H=1-\tilde\Pi^{\,\prime}_H=0$, 
where $\tilde\Pi^\prime_H$ is the derivative of the mass
function. The meson-mass functions are defined as 

\begin{eqnarray}
\Pi_P(p^2)&=& \int\!\! \frac{d^4k}{(2\pi)^4i} \tilde\Phi^2_P(-k^2) {\rm
tr} \biggl[\gamma^5 S_3(\not\! k) \gamma^5 S_1(\not\! k+\not\!
p)\biggr] \,, \label{massp}\\
\Pi_V(p^2)&=&\frac{1}{3}\biggl[g^{\mu\nu}-\frac{p^\mu
p^\nu}{p^2}\biggr] \int\!\! \frac{d^4k}{(2\pi)^4i} \tilde\Phi^2_V(-k^2)
{\rm tr} \biggl[\gamma^\mu S_3(\not\! k) \gamma^\nu S_1(\not\!
k+\not\! p)\biggr] \,.\label{massv}
\end{eqnarray}
In the baryon case the compositeness condition may be rewritten 
in a form suitable for the determination of the coupling constants:
\begin{eqnarray}
\hspace*{-.8cm}&&-12\, g^2_{q_1q_2q_3}
\int\!\frac{d^4k_1}{(2\pi)^4i}\!\int\!\frac{d^4k_2}{(2\pi)^4i}\,
\tilde\Phi^2_{q_1q_2q_3}(-k^2)\,D^\mu_{q_1q_2q_3}\,|_{\,\not p=m_H}\,
=\,\gamma^\mu\,, \label{coupl}\\
\hspace*{-.8cm}&&D^\mu_{q_1q_2q_3}=\gamma^\alpha\gamma^5\,
S_{q_1}(k_1+p)\,\gamma^\mu\,S_{q_1}(k_1+p)\gamma^\beta\gamma^5\,
{\rm tr}\left(
S_{q_2}(k_1+k_2)\,\gamma^\alpha\, S_{q_3}(k_2)\,\gamma^\beta\right)\,,
\nonumber\\
\hspace*{-.8cm}&&(q_1q_2q_3)=(bcq)\,,(bsq)\,,(csq)\,,\nonumber\\
\hspace*{-.8cm}&&\nonumber\\
\hspace*{-.8cm}
&&D^\mu_{ccq}=\gamma^\alpha\gamma^5\, S_{q}(k_2)\,\gamma^\beta\gamma^5\,
{\rm tr}\left(\gamma^\alpha\,
S_{c}(k_1+p)\,\gamma^\mu\,S_{c}(k_1+p)\,\gamma^\beta\,
S_{c}(k_1+k_2)\right)\,. \nonumber
\end{eqnarray}
The calculational techniques are outlined in Refs.~\cite{Bc,dhb}. 

Before presenting our numerical results we need to specify our values
for the constituent quark masses and shapes of the vertex functions.
As concerns the vertex functions, we found a good description of various
physical quantities \cite{Bc,model,RTQM} adopting a Gaussian form.
Here we apply the same procedure using 
$
\tilde\Phi_H(k^2_E)=\exp\{-k^2_E/\Lambda_H^2\}
$
in the Euclidean region. The magnitude of $\Lambda_H$ characterizes the
size of the vertex function and is an adjustable parameter in our model.
The $\Lambda_H$ parameters in the meson sector were determined \cite{Bc}
by a least-squares fit to experimental data and lattice determinations.
The quality of the fit may be seen from Table 1 for the leptonic
decay constants.
The nucleon $\Lambda_N$ parameter was determined from the best description of 
the electromagnetic properties of the nucleon \cite{model}. 
The $\Lambda_H$ parameters for baryons with one heavy quark (bottom or charm)
are determined by analyzing available experimental data on bottom and charm 
baryon  decays \cite{RTQM}. Since there is no  experimental information on 
the properties  of doubly heavy baryons yet we use the simple observation 
that the magnitude 
of $\Lambda_H$ is increasing  with the mass value of the hadron whose shape 
it determines. Keeping in mind that $\Lambda_N=1.25$ GeV, 
$\Lambda_{Qqq}=1.8$ GeV and $\Lambda_{B_c}=2.43$ GeV, we simply choose the 
value of $\Lambda_{QQq}=2.5$ GeV for the time being. We found that variations 
of this value by 10 $\%$  does not much affect the values of form factors. 
We employ the same values for the quark masses (see, Eq.(\ref{quarks})) 
as have been used previously for the description of light and heavy 
mesons (baryons) \cite{model,RTQM}. Note that the values for the light
quark masses in the meson case are different than in the baryon case
as a result of the lack of confinement in our approach. 

We thus use (in GeV)
\begin{equation}\label{quarks}
\begin{array}{cccc}
m_u & m_s & m_c & m_b \\
\hline
$\ \ (0.235) 0.420\ \ $ & $\ \ (0.333) 0.570\ \ $ &  $\ \ 1.67\ \ $ & 
$\ \ 5.06\ \ $  \\
\end{array}
\end{equation}

\begin{equation}\label{fitlam}
\begin{array}{ccccccccc}
\Lambda_\pi & \Lambda_K & \Lambda_D & \Lambda_{D_s} &
\Lambda_{J/\psi} & \Lambda_B & \Lambda_{B_s} & \Lambda_{B_c}&
\Lambda_{\Upsilon }\\ \hline $\ \ 1.16\ \ $ & $\ \ 1.82\ \ $ & $\
\ 1.87\ \ $ & $\ \ 1.95\ \ $ & $\ \ 2.12\ \ $ & $\ \ 2.16\ \ $ &
$\ \ 2.27\ \ $ &$\ \ 2.43\ \ $  & $\ \ 4.425\ \ $  \\
\end{array}
\end{equation}

The resulting form factors are {\it approximated} by the interpolating form
\begin{equation}\label{approx}
f(q^2)=\frac{f(0)}{1-a_1 q^2+a_2 q^4}
\end{equation}
It is interesting that for most of the form factors the numerical fit  
values of $a_1$ and $a_2$ obtained from the interpolating form~(\ref{approx}) 
are such that the form factors can be represented by monopole function
in the case of $B_c$-meson and dipole formula in the case of doubly
heavy baryons:
\begin{equation}\label{dipole}
f(q^2)\approx \frac{f(0)}{(1-q^2/m^2_V)^n} \hspace{0.5cm} (n=0,1).
\end{equation}
The values of $m_V$ in this representation are very close to the values 
of the appropriate lower-lying $(\bar q q')$ vector mesons 
($m_{D^\ast_s}$=2.11 GeV for (c-s)-transitions and 
$m_{B_c^\ast}\approx m_{B_c}$=6.4 GeV for (b-c)-transitions). In Fig.1 
we show two representative form factors and their dipole approximations. 
It is gratifying to see  that our relativistic quark model with the Gaussian 
vertex function and free quark propagators reproduces the monopole in the 
meson case and the dipole in the baryon case for most of the form factors. 

Finally, in Table 2  we present our predictions for the branching ratios
of the semileptonic $B_c$-decay rates and compare them with other approaches. 
In Table 3 the predictions for the decay widths of the doubly heavy
$\Xi$-baryons are given. We compare them with the free quark decay widths. 
One notes that the rates for the exclusive modes $\Xi_{bc}\to\Xi_{cc}+l\bar\nu$
and $\Xi_{bc}\to\Xi_{bs}+l\bar\nu$ are rather small when compared
to the total semileptonic inclusive rate estimated. The remaining
part of the inclusive rate would have to be filled in by decays into excited
or multi-body baryonic states. 
Note that the smallness of the exclusive/inclusive
ratio of the above exclusive modes
markedly differs from that of the mesonic semileptonic
$b\to c$ transitions, where the exclusive transitions to the ground state 
S-wave mesons $B\to D,D^\ast$ make up approximately 66$\%$ of the total 
semileptonic $B\to X_c$ rate \cite{PDG}. For $\Lambda_b\to\Lambda_c$ 
transitions one expects even higher semileptonic exclusive-inclusive ratio 
of amount 80$\%$ \cite{KM}. Note that the rate for 
$\Xi_{bc}\to\Xi_{cc}+l\bar\nu$ is of 
the same order of magnitude as the rates calculated for the corresponding 
double heavy mesonic decays $B_c\to\eta_c+l\bar\nu$ and 
$B_c\to J/\Psi+l\bar\nu$ \cite{Bc}. The QCD sum rule and potential model 
predictions for the rates of $\Xi_{bc}\to\Xi_{cc}+l\bar\nu$ and 
$\Xi_{bc}\to\Xi_{cs}+l\bar\nu$ given in \cite{Likh} exceed our rate 
predictions by factors of 10 and 3 respectively. In fact, the exclusive 
semileptonic rates given in \cite{Likh} tend to saturate the inclusive 
semileptonic rates.

In Table 4 we present  values for the invariant form factors at 
$q^2_{\rm min}=0$ and $q^2_{\rm max}=(m_i-m_f)^2$. Note that the values of 
the axial vector form factor $F_1^A$ are rather small for the two decays 
$\Xi_{bc}\to\Xi_{cc}+l\bar\nu$ and $\Xi_{bc}\to\Xi_{bs}+l\bar\nu$. 
This provides for a partial explanation of why the rates of these two modes 
are small compared to the inclusive semileptonic rate. Also the zero recoil 
values of the vector form factors are significantly below the value of one 
which one would expect from a naive application of the heavy quark limit. 
The smallness of the vector form factors provide for the remaining 
explanation of the smallness of the predicted respective rates. We mention 
that the QCD sum rule and potential model estimates of the zero recoil values 
of both the vector and axial vector form factors $F_1^V$ and $F_1^A$ given
in \cite{Likh} are close to one. In the model of \cite{Likh} the form factors 
$F_2^V$ and $F_2^A$ are set to zero. In our approach we find that the 
numerical values of $F_2^V$ and $F_2^A$ are quite small compared to those 
of $F_1^V$ and $F_1^A$ in all cases when expressed in terms of the mass 
scale $(m_i+m_f)$.

\newpage

\begin{table}[t1]
\caption {Leptonic decay constants $f_H$ (MeV) used in the least-squares fit.}
\begin{center}
\begin{tabular}{|l|l|l|}
\hline Meson & This model & Expt./Lattice  \\
\hline\hline
   $\pi^+$ & 131 & $130.7\pm 0.1\pm 0.36$       \\
   $K^+$   & 160 & $159.8\pm 1.4\pm 0.44$      \\
   $D^+$   & 191 & $191^{+19}_{-28}$   \\
   $\,\, D^+_s$ & 206 & $206^{+18}_{-28}$      \\   
   $B^+$   & 172 & $172^{+27}_{-31}$   \\
$\,\, B^+_s$& 196   & $171\pm 10^{+34+27}_{-9-2} $  \\
 $\,\, B_c$  & 360  & 360       \\ 
\hline 
$J/\psi$ &404   & 405$\pm$ 17 \\
$\Upsilon$ & 711 & 710$\pm$37 \\
\hline\hline  
\end{tabular}
\end{center}
\end{table}

\begin{table}[t2]
\caption{ Branching ratios BR($\%$) for the  semileptonic decays
$B_c^+ \rightarrow H l^+ \protect\nu $, calculated with
the CDF central value $\protect\tau (B_c)=0.46$ ps $\protect\cite{CDF}$.}
\begin{center}
\begin{tabular}{|c|c|c|c|c|c|c|c|}
\hline\hline H & This model & \cite{KLO}
&\cite{AMV}&\cite{CC}&\cite{CF1}&\cite{AKNT} &\cite{NW}\\
\hline\hline
$\eta_c\, e\,\nu$ &0.98 & 0.8$\pm$0.1 & 0.78& 1.0 & 0.15(0.5) & 0.6 & 0.52 \\
\hline
$\eta_c\,\tau\,\nu$   & 0.27   & & & &  &  &  \\
\hline
$ J/\psi\, e\,\nu$   & 2.30  &
2.1$\pm$0.4 & 2.11 & 2.4 & 1.5(3.3) & 1.2 & 1.47\\
\hline
$J/\psi \,\tau\,\nu$    & 0.59   & &  & &  &  &  \\
\hline
$ D^0\, e\,\nu$  & 0.018  &     & 0.003 & 0.006 & 0.0003(0.002) & &  \\
\hline
$ D^0\,\tau\,\nu$    & 0.0094  &  & &  &  & & \\
\hline
$ D^{\ast 0} \, e\,\nu$ & 0.034  &      & 0.013 & 0.019 &0.008(0.03) & &  \\
$D^{\ast 0} \,\tau\,\nu$    & 0.019 &  & &  &  & & \\
\hline\hline
$B^0\, e\,\nu $     &0.15  &       & 0.08 & 0.16 & 0.06(0.07) & & \\
\hline
$ B^{\ast 0}\, e\,\nu$   & 0.16  &     &0.25 & 0.23 &0.19(0.22)  & &    \\
\hline
$ B^0_s\, e\,\nu $    & 2.00  & 4.0 & 1.0 & 1.86 & 0.8(0.9) & 1.0 & 0.94 \\
\hline $ B_s^{\ast 0}\, e\,\nu$ & 2.6   & 5.0  & 3.52 & 3.07  &2.3(2.5)  & & 1.44 \\
\hline
\end{tabular}
\end{center}
\end{table}

\begin{table}[t3]
\caption{Calculated decay widths of lowest lying $J^P=1/2^+$ doubly heavy 
$\Xi$-baryons. Inclusive widths are calculated using the current
quark pole masses.}
\vspace*{0.2cm}
\def\arraystretch{1.85}
\begin{center}
\begin{tabular}{|c|c|r|}
\hline
    & \multicolumn{2}{|c|} {Decay widths, ps$^{-1}$} \\
\cline{2-3} Mode  & RTQM & Inclusive width  \\
\hline\hline
$\Xi_{\,\rm bc}\,\to\,\Xi_{\,\rm cc}\,+\, l\,\bar\nu$
& \,0.012\, & 
 $2\cdot\Gamma_0(b\to c)$\,=\,0.162 \\
\hline
$\Xi_{\,\rm bc}\,\to\, \Xi_{\,\rm bs}\,+\, l\,\bar\nu$
& \,0.043 \, & 
$\Gamma_0(c\to s)$\,=\, 0.122\\
\hline
$\Xi_{\,\rm cc}\,\to\, \Xi_{\,\rm cs}\,+\, l\,\bar\nu$
& \,0.224\,  &
$2\cdot\Gamma_0(c\to s)$\,=\,0.244\,\\
\hline 
\end{tabular}
\end{center}
\end{table}

\begin{table}[t4]
\caption{Values of $F_1^V$ and $F_1^A$ form factors at maximum and zero 
recoil.}
\vspace*{0.2cm}
\def\arraystretch{1.85}
\begin{center}
\begin{tabular}{|l|cc|cc|cc|}
\hline
     & \multicolumn{2}{|c|} {$\Xi_{bc}\to \Xi_{cc}$} & 
       \multicolumn{2}{|c|} {$\Xi_{bc}\to \Xi_{bs}$} & 
       \multicolumn{2}{|c|} {$\Xi_{cc}\to\Xi_{cs}$}  \\
\cline{2-7} & $F^V_1$ & $F^A_1$ & $F^V_1$ & $F^A_1$  & $F^V_1$ & $F^A_1$  \\
\hline\hline
$q^2=0$ &\, 0.46\,  &\, -0.091\, &\, 0.39\, &
\, 0.061 \,& \,0.47\, &\, 0.61\, \\
\hline
$q^2=q^2_{\rm max}$ &\, 0.83\, &\, -0.086\, &\, 0.58\, &\, 0.065 \,& 
\, 0.59\, &\, 0.77\, \\ 
\hline 
\end{tabular}
\end{center}
\end{table}

\begin{figure}[t]
\begin{center}
\begin{tabular}{c}
\epsfig{figure=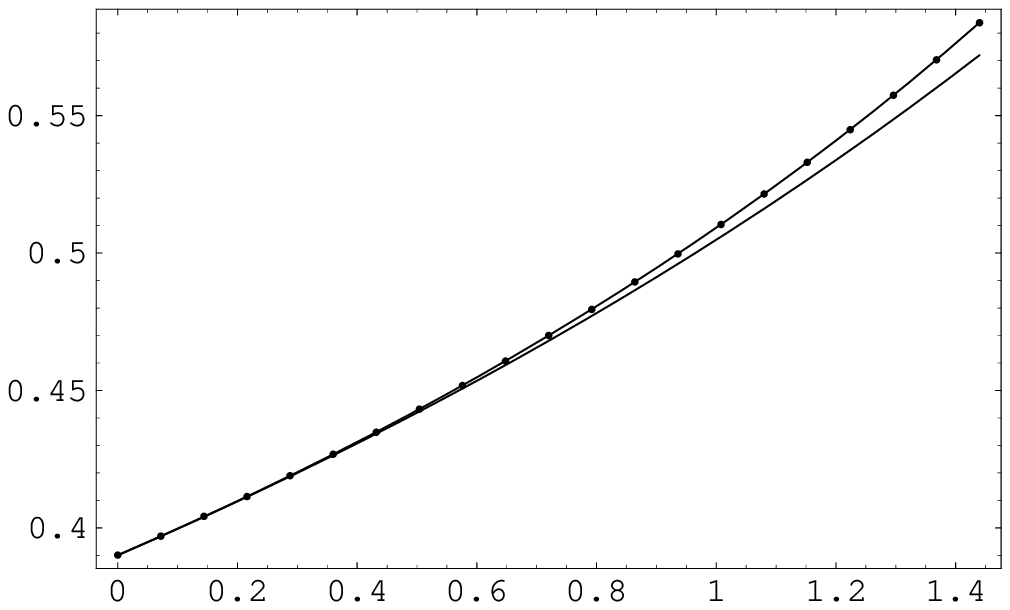,height=8.3cm}\\
\epsfig{figure=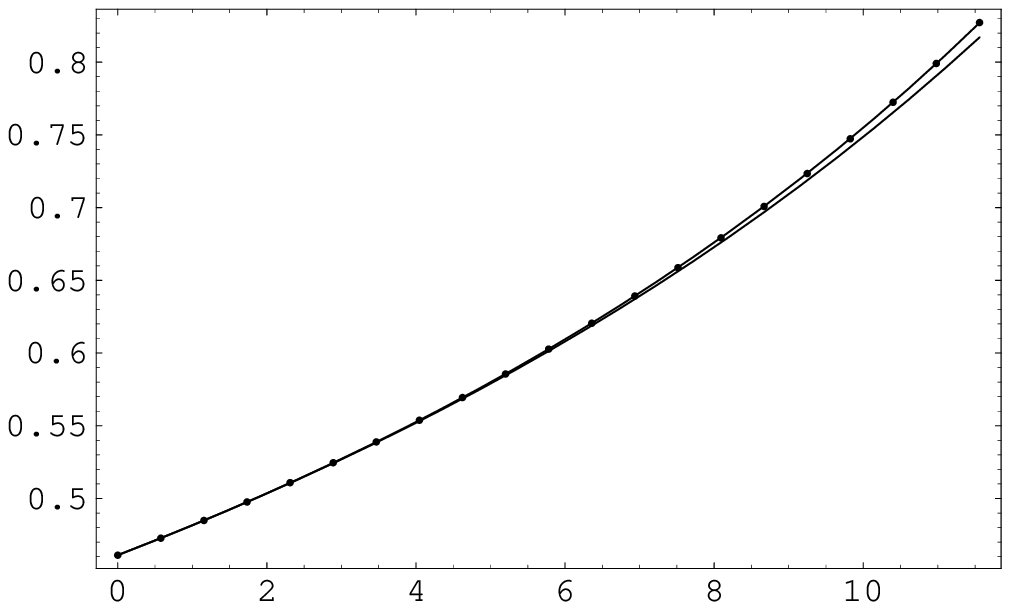,height=8.3cm}
\end{tabular}
\end{center}
\caption{
Upper panel:  Form factor $F_1^V (q^2)$ (solid-dotted line)
for $bc\to bs$ transitions and its dipole 
approximation (solid line) with $m_{cs}=2.88$ GeV;
Lower panel:  Form factor $F_1^V (q^2)$ (solid-dotted line)
for  $bc\to cc$ transitions and its dipole 
approximation (solid line) with  $m_{bc}=6.81$ GeV.
}
\end{figure}

\end{document}